# The Web is missing an essential part of infrastructure: an Open Web Index

Dirk Lewandowski
Hamburg University of Applied Sciences, Hamburg, Germany
dirk.lewandowski@haw-hamburg.de

*A proposal for building an index of the Web that separates the infrastructure part of the search engine—the index—from the services part that will form the basis for myriad search engines and other services utilizing Web data on top of a public infrastructure open to everyone*

The Web as we know it would not be possible without search engines. They are an integral part of the Web and can also be seen as a part of the Web's infrastructure. Google alone now serves over 2,000,000,000,000 search queries per year [11]. While there seem to be a multitude of search engines on the market, there are only a few relevant search engines in terms of them having their own index (the database of web pages underlying a search engine). Other search engines pull results from one of these search engines (e.g., Yahoo pulls results from Bing), and should therefore not be considered search engines in the true sense of the word. Globally, the major search engines with their own indexes are Google, Bing, Yandex and Baidu. Other independent search engines may have their own indexes, but not to the extent that their size makes them competitive on the global search engine market.

While the search engine market in the U.S. is split between Google and Bing (and its partner Yahoo) with roughly two thirds to one-third, respectively [10], in most European countries, Google accounts for more than 90 percent of the market share. As this situation has been stable over at least the last few years, there have been discussions about how much power Google has over what users get to see from the Web, as well as about anti-competitive business practices, most notably in the context of the European Commission's competitive investigation into the search giant [3].

**Search engine bias?**

From the users' point of view, search engines are reliable and trustworthy sources, providing fair and unbiased results [8]. However, it has been found that search results simply should not be considered "neutral". Some scholars argue that an unbiased search engine is simply not possible, as there is no ideal result set against which a bias can be measured [5,6]. Therefore, I argue that every search engine presents its own algorithmically generated view of the web's content. Every such view can be different, and none of them are the definitive or correct one.

Problems that may arise from search engines' interpreting the world in certain ways are, amongst others, reinforcing stereotypes, e.g., towards women [7]; influencing public opinion in the context of political elections (e.g., [2]), and preferring dramatic interpretations of rather harmless health-related symptoms [13].

It seems, therefore, unreasonable to have only one (or a few) dominant search engines imposing their view on the Web's content, which is, on closer inspection, really only one of many possible views. Therefore, I argue for building an index of the Web that will form the basis for a multitude of search engines and other services that are based on Web data.

**Three major problems**

There are three major problems resulting from a search engine market where only a few competitors are equipped with their own index of web pages:

(1) A search engine provides only one of many possible algorithmic interpretations of the Web's content. At least for informational queries (cf. [1]), there is no correct set of results, let alone one single correct result. For these queries, we usually find a multitude of results of comparable quality. While a search engine's ranking might provide some relevant results on the highest positions, there may be many more (or to some users, even better) results on lower positions.

(2) Every search engine faces a conflict of interest when it also acts as a content provider and shows results from its own offerings on its results pages (e.g., Google showing results from its subsidiary YouTube). This problem gets exacerbated when one search engine has a large market share, as it is able to increase both its influence on its users as well as its suppression of its competitors' offerings.

(3) The more users rely on a single search engine, the higher the influence of search engine optimization (SEO) on the search results, and therefore, on what users get to see from the Web. The aim of SEO is to optimize web pages so that they get ranked higher in search engines (i.e., influencing a search engine's results). Taken together with the fact that SEO is now a multi-billion dollar industry [12], we can see huge external influences on search engine results.

**A lack of plurality**

Considering these three problems, we can see that in the current market situation, we are far from plurality, not only in terms of the numbers of search engine providers but also in the number of search results. In 2011, a study from Yahoo showed that while we can regard a search engine as a possible window to all of the web's content, more that 80% of all user clicks were found to go to only 10,000 different domains [4]. We can assume that these numbers are comparable for other search engines. Taken together, search engines have a huge influence on what we as users get to see on the results pages, and consequently, what we select from.

**Why are there no alternatives?**

Why are there no real alternatives to the few popular search engine index providers? Firstly, index providers face huge technical difficulties due to the large numbers of documents resulting from the ever-changing nature of the Web. A second, significant, issue is the cost of hardware, infrastructure, maintenance and staff. Thirdly, the Web is huge, and a search engine index needs to be tasked with covering as large a part of it as possible. While we know that no search engines can cover the Web in total, modern search engines know of trillions of existing pages [9]. And indexing these pages is only the start. A search engine needs to keep its index current, meaning it needs to update at least a part of it every minute. This is an important requirement that is not being met by any of the current projects (like *Common Crawl*) aiming at indexing snapshots of (parts of) the Web.

**Separating index and services**

I am proposing an idea for a missing part of the Web's infrastructure, namely a searchable index. The idea is to separate the infrastructure part of the search engine (the index) from the services part, thereby allowing for a multitude of services, whether existing as search engines or otherwise, to be run on a shared infrastructure (see Fig. 1).

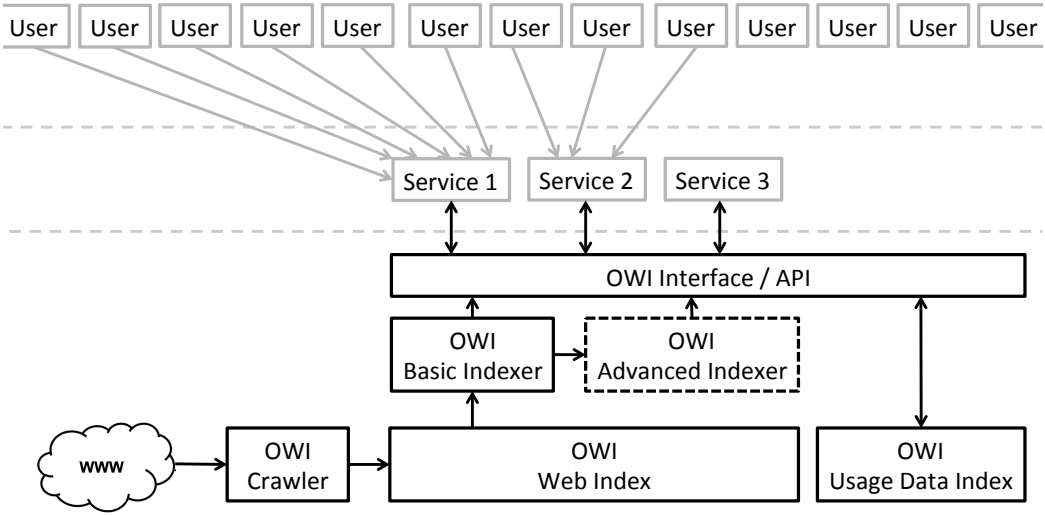

Fig.1: Separating services from infrastructure

The figure shows how the public infrastructure is responsible for crawling the web, for indexing its content, and for providing an interface/API to the services that are built upon the index.

The indexing stage is divided between basic indexing and advanced indexing. Basic indexing provides the data in a form that services built on top of the index can easily and rapidly process that data. So, while services are allowed to do their further indexing to prepare documents, some advanced indexing is also provided by the open infrastructure. This provides additional information to the indexed documents (e.g., semantic annotations). For this, an extensive infrastructure for data mining and processing is needed. Services should, however, be able to decide for themselves to what extent they want to rely on the pre-processing infrastructure provided by the Open Web Index. A design principle should be to allow services a maximum of flexibility.

As modern search engines rely heavily on usage data, these data (most prominently search queries routed to the index) are collected and made available for reuse. The OWI Usage Data Index allows for this data to be collected, stored and queried. So, while each service can collect and query their own usage data, every service that wants to access usage data from the OWI Usage Data Index should be required to share anonymized usage data with the other services, so that every service profits from the amassed data. It is clear that existing search engines like Google and Bing have a huge lead compared to new providers, as they have a solid user base and already amassed lots usage data. However, sharing usage data between the services could at least lessen the cold start problem.

**Benefits**

The main benefit of such an index would be for all interested parties to be able to develop their own applications without the problem of having to create their own index of the Web, which currently is an impossible endeavor not only, but especially, for small and medium enterprises, as well as for non-commercial bodies.

Given a considerable uptake for such an index, it would foster plurality not only in the use of Web content by developers but also in the variety of content that users get to see. We can rightly assume that each search engine using the index would apply its own ranking function, and therefore, produce different results. Users would benefit in that they would not have to rely on only one or at best a few search engines but could choose from a variety of engines serving their different purposes. In that way, an open web index would foster plurality and restrict the power of single companies dictating which content is shown to and consumed by users.

Another benefit would be that the index would be open to everyone, and therefore, would allow for investigating its transparency. However, search engines built on top of the index could still be "black boxes" in that they would not need to make their ranking functions open to anybody.

**Possible applications**

While the Open Web Index would first and foremost make the development of new Web search engines feasible and financially attractive, it could also form the basis for a variety of other applications, being related to search or not.

In the field of search, the Open Web Index would also allow for vertical search engines (like image search, video search or search in specific areas and on specific topics) to be built. In vertical search applications, OWI data could also be used to amend proprietary data. For instance, a provider of company information could amend its company profiles with web data.

Apart from search, the OWI could also build the basis for data analysis and topic detection and tracking. Examples of applications are opinion mining tools and market research applications.

In the field of artificial intelligence, the Open Web Index could be used as a basis for large-scale machine learning. Likely applications in this area are machine translation, question answering and conversational applications.

Last but not least, an Open Web Index would provide a rich data source for researchers in many different fields, ranging from computer science and computational linguistics to computational social sciences and research evaluation.

It is clear that this short list of ideas is far from being complete and only serves illustrative purposes. It shows, however, the huge potential of making web data open to all parties interested.

**Alternative approaches**

Some alternative solutions have been proposed for fostering plurality on the search engine market. The first and probably obvious solution is to wait for commercial market players to develop alternatives. However, as we have seen in the last fifteen years or so, Bing has been the only search engine capable of gaining considerable market share. Other search engines have failed, have been acquired by larger search companies or have focused on niche markets. All new search engine providers face the problem of having to build their own index, which is, as has been described earlier, a very costly undertaking. Furthermore, what would be gained if we had one or two, even three more search engines on the market? From my point of view, the problem lies not in having a few more search engines, but in providing real search plurality.

The second line of argumentation says that Google should be forced to provide fair and unbiased results. This is what the European Commission's competitive investigation against Google has been all about. However, as ranking results is always based on interpretations (and human assumptions inherent in the ranking algorithms), there is no such thing as an unbiased result set. Only a multitude of different algorithmic interpretations can help bring about search plurality.

The third line of argumentation calls for Google to open its index to third parties. Then, it would be possible to build (search) applications on top of Google's index. However, the control over the index – and over what third parties would be able to get from the index – would still lie in the hands of a private company, the index would still not be transparent, and there would still be no influence on how the index is composed.

The fourth, and already widely discussed solution, is building a publicly funded search engine as an alternative to the commercial enterprises. However, this again would only add one more search engine to the market, instead of fostering plurality.

**Conclusion**

The main idea I presented is to foster building search engines and other services needing Web data on top of a public infrastructure that is open to everyone. A multitude of such services would foster plurality not only on the search engine market (with the result of having more than a few search engines to choose from) but even more importantly, a plurality with regards to the results users get to see when using search engines.

Search results as a basis for knowledge acquisition in society seem too important to be left solely in the hands of a few commercial enterprises. The Open Web Index is comparable to other public services such as constructing roads and railroad tracks, supporting public broadcasting and, most notably, building a library system. An Open Web Index could be one of the main building blocks of the library of the 21$^{st}$ century. An open web index is a project that cannot and should not be undertaken by a single company or institution. On the contrary, I see building such an index as a task of society and for society, meaning that we should build the index involving all actors and interest groups relevant to society at large. Those that benefit from the index should have their say in building it.

A question that remains is funding. As a considerable amount of money is needed, I argue for public funding not by a single state, but rather by a larger entity like the European Union. This should, however, not mean that a governmental body should also be the operator of the Open Web Index. Rather, it should be run by an organization that is relatively free from state intervention. One could think of a foundation running it or a model similar to public broadcasting. Whatever the mode of operation, as a project of and for society, funding should be applied for the greater good.

**REFERENCES**

1. Broder, A. A taxonomy of web search. *ACM Sigir forum 36*, 2 (2002), 3–10.
2. Epstein, R. und Robertson, R.E. The search engine manipulation effect (SEME) and its possible impact on the outcomes of elections. *Proceedings of the National Academy of Sciences 112*, 33 (2015), E4512–E4521.
3. European Commission. Antitrust: Commission fines Google €2.42 billion for abusing dominance as search engine by giving illegal advantage to own comparison shopping service - Factsheet. 2017. http://europa.eu/rapid/press-release_MEMO-17-1785_en.htm.
4. Goel, S., Broder, A., Gabrilovich, E., und Pang, B. Anatomy of the long tail: Ordinary people with extraordinary tastes. *Proceedings of the third ACM international conference on Web search and data mining*, ACM (2010), 201–210.
5. Grimmelmann, J. Some Skepticism About Search Neutrality. *The next digital decade: Essays on the future of the internet 31*, (2010), 435–460.

6. Lewandowski, D. Is Google Responsible for Providing Fair and Unbiased Results? In M. Taddeo und L. Floridi, Hrsg., *The Responsibilities of Online Service Providers*. Springer, Berlin Heidelberg, 2017, 61–77.
7. Noble, S.U. Google Search: Hyper-visibility as a Means of Rendering Black Women and Girls Invisible. *InVisible Culture: An Electronic Journal for Visual Culture*, 2013. http://ivc.lib.rochester.edu/google-search-hyper-visibility-as-a-means-of-rendering-black-women-and-girls-invisible/.
8. Purcell, K., Brenner, J., und Raine, L. *Search Engine Use 2012*. Washington, DC, 2012.
9. Schwartz, B. Google's search knows about over 130 trillion pages. *Search Engine Land*, 2016. http://searchengineland.com/googles-search-indexes-hits-130-trillion-pages-documents-263378.
10. Sterling, G. Data: Google monthly search volume dwarfs rivals because of mobile advantage. *Search Engine Land*, 2017.
11. Sullivan, D. Google now handles at least 2 trillion searches per year. *Search Engine Land*, 2016. http://searchengineland.com/google-now-handles-2-999-trillion-searches-per-year-250247.
12. Sullivan, L. Report: Companies Will Spend $65 Billion On SEO In 2016. *Media Post*, 2016. http://www.mediapost.com/publications/article/273956/report-companies-will-spend-65-billion-on-seo-in.html.
13. White, R.W. und Horvitz, E. Cyberchondria. *ACM Transactions on Information Systems 27*, 4 (2009), Article No. 23.